\documentclass[a4paper,twoside]{article}
\newcommand{\affil}[1]{$^{\rm #1}$}
\usepackage[authoryear]{natbib}
\bibpunct{(}{)}{;}{a}{}{,}
\usepackage{graphicx}
\date{} 
\title{\large\bf\flushleft High-Level Magnetic Activity on Low Mass Close Binary: \\
GSC 2038-0293}
\author{\parbox{\textwidth}{\flushleft
\vspace{-0.5cm}
{\it DAL, H.A. \affil{1},\affil{2}, S\.{I}PAH\.{I}, E. \affil{1}, \"{O}ZDARCAN}, O. \affil{1} \\
\vspace{0.4cm}
{\small \affil{1}\,Department of Astronomy and Space Sciences, University of Ege, Bornova, 35100 ~\.{I}zmir, Turkey}\\
{\small \affil{2}\,Corresponding Author, Email: ali.dal@ege.edu.tr}}}
\begin{document}
\twocolumn[
\begin{changemargin}{.8cm}{.5cm}
\begin{minipage}{.9\textwidth}
\vspace{-1cm}
\maketitle
%
%
\small{\bf Abstract:} Taking into account results obtained from light curve analysis and from analyses of out-of-eclipses, we discussed the nature of GSC 02038-00293 and also its magnetic activity behaviour. We obtained the light curves of the system in observing seasons 2007, 2008 and 2011. We obtained its secondary minimum clearly in I band observations in 2008 for the first time. Analysing this light curve, we found physical parameters of the components. The light curve analysis indicates that possible mass ratio of the system is 0.35. We obtained the remained V band light curves, extracting the eclipses. We modelled these remained curves by using the SPOTMODEL program, and we found possible spot configurations of magnetically active component for each observing season. The models demonstrated that there are two active longitudes on the active component. The models reveal that both active longitudes migrate toward the decreasing longitudes. We also examined the light curves at out-of-eclipses in respect to its minimum and maximum brightness and the amplitude, and etc. The amplitude of the curves at out-of-eclipses is varying in a sinusoidal way with a period of $\sim$8.9 years; the mean brightness of the system is dramatically decreasing. The phases of the deeper minimum at out-of-eclipses exhibit a migration toward the decreasing phases.\\


\medskip
\medskip
\end{minipage}
\end{changemargin}
]
\small

\section{Introduction}

Many stars such as BY Dra and RS CVn type stars on the main sequence toward the late spectral types exhibit stellar spot activity called as magnetic activity. In the literature, the BY Dra variables were found by \citet{Kro52} for the first time, who demonstrated some sinusoidal-like variations at out-of-eclipses of the eclipsing binary star YY Gem. He explained the variation at out-of-eclipses as a heterogeneous temperature on star surface. Then, it was called BY Dra stars by \citet{Kun75}. Based on some rigorous arguments obtained from observations, the variability seen in BY Dra type stars was confirmed in terms of dark regions on the surface of the stars by later works such as \citet{Tor73, Bop73, Vog75, Fri75, Bop77}. RS Cvn type stars exhibit the same variability with BY Dra stars. However, one of the components of RS CVn stars is evolved and it is generally a giant or subgiant star, while the other component is a main sequence star \citep{Tho08}. The incidence of late type stars in our Galaxy is about 65 $\%$. Seventy-five percent of them show magnetic activity such as the spot and flare activities \citep{Rod86}. The researches such as \citet{Lop07, Mor08, Mor10} demonstrate that magnetic activity dramatically affects the stellar structure of the late type stars and also their evolutions. \citet{Mor08, Cas08, Fer09, Mor10, Tor10, Kra11} have revealed that the radii found from the analyses of the observations are generally larger than the radii theoretically expected for several magnetically active low mass binaries, while the effective temperatures found from the observations are usually lower than those theoretically expected. They claimed that the reason is most likely magnetic activity. In this respect, magnetically active low mass components of the binaries take very important places to understand their evolution.

In this study, we introduce GSC 02038-00293 as a new candidate for the magnetically active low mass stars. Using the ROTSE 1 database \citep{Woz04}, the system was discovered by \citet{Ber06} in the optical identification program of X-ray sources listed in the ROSAT All-Sky Survey Bright Source Catalogue \citep{Vog99}. The identification reveals that the system is actually the uncatalogued variable NSVS object ID 7869362 as the optic counterpart of the X-ray source 1RXS J160248.3+252031. Combining their 2005 and 2006 data with the available data taken from the ROTSE 1 and ASAS 3 databases, \citet{Ber06} determined light elements as follows: $JD(Hel.)_{Min I}$ = 53560.491(3)+$0^{d}.49541(1)$ $\times E$. \citet{Nor07} gave the same period analysing the Super-WASP observations of the system. In addition, \citet{Fra07} also confirmed the period with 2007 observations. \citet{Ber06} indicted a 6 - 8 year long activity cycle. However, consecutive observations demonstrated that the light curve shapes can change even in one week. Finally, \citet{Ber06} identified the system as a RS CVn type binary. Using low resolution spectra, Dragomir et al. (2007) indicated that the spectral type of the system likely to be a K type. A detailed spectral study was done by \citet{Kor10}. Considering some features seen in the low resolution spectra of the system such as neutral metals ($Mg$, $Na$), weak Balmer lines and also absence of molecular bands, they confirmed that the system is from K spectral types. Adopting $log~g=4.5$, the $T_{eff}$ was found to be $4750\pm250$ $K$ and the $v \sin i$ was found to be $90\pm10$ $kms^{-1}$. However, using their the v sin i measured together with the rotation period of 0.495410 days found by \citet{Ber06}, \citet{Kor10} estimated the system's radius as $R \times \sin i = 0.88\pm0.10$ $R_{\odot}$. The value of $R \times \sin i$ indicated a late-G spectral type or later.

The observations in the literature demonstrated that GSC 02038-00293 exhibits magnetic activity. In this study we analysed the light curves of the system, and also we examined the variations at out-of-eclipses. We finally compared the system with its analogue in respect to the theoretical models.

\section{Observations and Data}

Observations of GSC 02038-00293 were carried on with two telescopes in BVRI bands at Ege University Observatory. The first part of observations was acquired with a High-Speed Three Channel Photometer attached to the 48 cm Cassegrain type telescope in observing seasons 2007 and 2008. The second part of observations was acquired with a thermoelectrically cooled ALTA U+42 2048$\times$2048 pixel CCD camera attached to a 40 cm - Schmidt - Cassegrains - type MEADE telescope in the observing seasons  2008 and 2011. The comparison and check stars used in all observations are the same stars used in the literature. Some basic parameters of program stars are listed in Table 1. The names of the stars are listed in first column, while J2000 coordinates are listed in second column. The V magnitudes are in third column, and B-V colours are listed in the last column.

Although the program and comparison stars are very close on the sky, differential atmospheric extinction corrections were applied. The atmospheric extinction coefficients were obtained from observations of the comparison stars on each night. Moreover, the comparison stars were observed with the standard stars in their vicinity and the reduced differential magnitudes, in the sense variable minus comparison, were transformed to the standard system using procedures outlined by \citet{Har62}. The standard stars are listed in catalogues of \citet{Lan83, Lan92}. Furthermore, the dereddened colours of the system were computed. Heliocentric corrections were also applied to the times of the observations.

In BVR bands, the first part of observations was continued for 6 nights between April 22 and July 19 in 2007, and it was carried on for 4 nights between April 7 and July 23 in 2008 with 48 cm Cassegrain type telescope. In addition, the second part of observations was continued for 9 nights between May 20, 2008 and August 20, 2008 in BVRI bands, and it was carried on for 2 nights on April 23 and May 9 in 2011 with 40 cm - Schmidt - Cassegrains - type MEADE telescope. The mean averages of the standard deviations were found to be $0^{m}.009$, $0^{m}.007$ and $0^{m}.007$ from the observations of 48 cm Cassegrain type telescope for BVR bands, respectively. They were found to be $0^{m}.023$, $0^{m}.011$, $0^{m}.010$ and $0^{m}.013$ for observations acquired with 40 cm - Schmidt - Cassegrains - type MEADE telescope in the BVRI bands, respectively. To compute the standard deviations of observations, we used the standard deviations of the reduced differential magnitudes in the sense comparison minus check stars for each night. There was no variation observed in the standard brightness of comparison stars.

The minima times obtained in this study are listed in Table 2. In the table, the first 9 minimum times have already been published by \citet{Sip09}, while the last one is unpublished. Using all available minima times in the literature, we adjusted the light elements of the system, as follows:

\begin{center}
\begin{equation}
JD~(Hel.)~=~24~53560.4925(9)~+~0^{d}.4954115(5)~\times~E.
\end{equation}
\end{center}

Using the light elements given by Equation (1), we phased all our observations and also all data taken from the literature. In Figure 1, the light and colour curves of the system are shown for three observing seasons 2007, 2008 and 2011. As it is seen from the figures, there is a remarkable variation in the shape of the light curves from a season to next one. As it is well known from the literature, the system exhibits magnetic activity. Figure 1 demonstrates that available magnetic activity causes dramatic distortion on the light curve shape. Although the secondary minimum can show itself in observing season 2008, it can not clearly reveal itself in general. In addition, the B-V colour curves exhibits some variations, while there is no clear variation over the standard deviation in the V-R colour curves. However, we have a chance to observe the system in I band in the program of the 40 cm - Schmidt - Cassegrains - type MEADE telescope for both seasons 2008 and 2011. I band observations show the secondary minimum better than all other bands (see the Section 3). The secondary minimum exhibits itself better toward long wavelengths.

In order to analyse, we collected all available data from the literature. For this aim, we got the ROTSE 1's V band data from the Northern Sky Variability Survey (hereafter NSVS) database \citep{Woz04}, and also we got the available data in the ASAS Database \citep{Poj97}. In addition, we took the observations published by \citet{Ber06} and \citet{Fra07}. The standard V band data of ROTSE 1 cover the observing seasons of 1999 and 2000, while the standard V band data of ASAS cover the seasons from 2003 to 2006. The observations in this study were started in 2007. Although the data taken from \citet{Ber06} and \citet{Fra07} were not standard, using the comparison and check stars given by them, we transformed to the standard system. After transforming their data, we compared all the available data whether the data taken from different sources are suitable to use together. For this aim, we compared the ASAS data of 2005 and 2006 with the data taken from \citet{Ber06}, and also we compared the data taken from \citet{Fra07} with our 2007 data. As seen from the comparisons, the levels of the data taken from different sources are statistically the same in 3$\sigma$ value. The light curves obtained from the available data are shown in Figure 2. As seen from the figure, the light variation shapes of seasons 1999, 2000, and 2005 are similar to the variation observed in season 2011. The light curves obtained in 2011 are similar to the light curves obtained previous studies. Fortunately, I band observations of 2008 gave a chance to us; we can do the light curve analyses. In the case of lack of the secondary minimum, the light curve analyse does not give reliable results.

\section{Light Curve Analysis}

The light curve analysis of such a magnetically active star is generally quite difficult due to the absence of the secondary minima. In fact, no secondary minimum is seen in our observations of the season 2007 and also generally all the other observations published in the literature. However, I band observations in 2008 clearly show the secondary minimum. This gave us a chance for the light curve analysis. This is why we analysed only I light curve obtained in 2008 with using the PHOEBE V.0.31a software \citep{Prv05}, whose method depends on the method used in the version 2003 of the Wilson-Devinney Code \citep{Wil71, Wil90}. We tried to analyse I band curve with three different modes, such as the "detached system", "semi-detached system with the primary component filling its Roche-Lobe" and "semi-detached system with the secondary component filling its Roche-Lobe" modes. The initial analyses demonstrated that an astrophysical acceptable result can be obtained if the analysis is carried out in the "detached system" mode. The initial experience revealed that no acceptable results in the astrophysical sense could be obtained in all the others modes.

Using low resolution spectra, \citet{Kor10} found that the system is from K spectral types, and the $T_{eff}$ was found to be $4750\pm250$ $K$. Considering the case, the temperature of the primary component was fixed to 4750 K, and the temperature of the secondary was taken as a free parameter in the analyses. Considering the spectral type, the albedos ($A_{1}$ and $A_{2}$) and the gravity darkening coefficients ($g_{1}$ and $g_{2}$) of the components were adopted for the stars with the convective envelopes \citep{Luc67, Ruc69}. The non-linear limb-darkening coefficients ($x_{1}$ and $x_{2}$) of the components were taken from \citet{Van93}. In the analyses, their dimensionless potentials ($\Omega_{1}$ and $\Omega_{2}$), the fractional luminosity ($L_{1}$) of the primary component and the inclination ($i$) of the system were taken as the adjustable free parameters.

There is no obtained spectroscopic mass ratio for the system. Because of this, we used the "q-search" method with using a step of 0.05 to find the best photometric mass ratio of the components. The general result of the q-search is shown in Figure 3. As seen from the figure, the minimum sum of weighted squared residuals ($\Sigma res^{2}$) is found for the mass ratio value of $q= 0.35$. According to this result, we assume that a possible mass ratio of the system is $q= 0.35$. 

As it is clearly seen from Figures 1, 2, there is a dramatic asymmetry in the light curves due to the magnetic activity. In the light curve analysis, we assumed that the primary component has two cool spots on its surface to remove this asymmetry. According to the results obtained from the first iterations, the secondary component seems to be a M dwarf, which is close to the full-convective boundary \citep{Bro11}. The full-convective M dwarfs exhibit very strong flare activity, while a few of them just exhibit spot activity \citep{Dal11}. In addition, if the secondary component is the spotted one, to remove the asymmetries seen in the light curves, the analysis demonstrated that the spots should be as large as that cover all the surface of the star or their effective temperatures must be half of the surface temperature due to the secondary component's light rate in the total light of the system. Although there are a few stars, which are close to the full-convective boundary and have some large spots on their surface. But, this is not a common phenomena. However, the K dwarfs are generally potential stars, which are possible to exhibit spot activity. This is why we assumed the spotted star is the primary component. Moreover, the light curve analyses with this assumption gave more acceptable results in the astrophysical sense. On the other hand, it must be noted here that it is well known that spot solution suffers from non-uniqueness. The synthetic light curve obtained from the light curve solution is seen in Figure 4, and the resulting parameters of the analysis are also listed in Table 3.

Although there is not any available radial velocity curve, we tried to estimate the absolute parameters of the components. According to \citet{Tok00}, the mass of the primary component must be 0.73$\pm$0.05 $M_{\odot}$ corresponding to its surface temperature. Considering possible mass ratio of the system, the mass of the secondary component was found to be 0.25$\pm$0.04 $M_{\odot}$.

Using Kepler's third law, we calculated possible the semi-major axis as a 2.62 $R_{\odot}$. Considering this estimated semi-major axis, the radius of the primary component was computed as 0.87$\pm$0.05 $R_{\odot}$, while it was computed as 0.27$\pm$0.04 $R_{\odot}$ for the secondary component. As it is seen from these results, the primary component's radius derived with the assumption of spotted primary component is in agreement with the value of $R \times \sin i$ found by \citet{Kor10}. Using the estimated radii and the obtained temperatures of the components, the luminosity of the primary component was estimated to be 0.35 $L_{\odot}$, and it was found as 0.01 $L_{\odot}$ for the secondary component. The absolute parameters are generally acceptable in the astrophysical sense. However, the radius of the primary component is larger than the expected values in respect to the theoretical models. We plotted the distribution of the radii versus the masses for some stars in Figure 5. The filled circles in the figure represent well-known active stars listed in the catalogue of \citet{Ger99}. Some of these stars exhibit the spot activity, while some of them exhibit the flare activity. Some stars exhibit both spot and flare activities. In the figure, the asterisk represents the secondary component, while the open triangle represents the primary component. The line represents the ZAMS theoretical model developed for the stars with $Z=0.02$ by \citet{Sie00}.

\section{Variations at Out-of-Eclipses}

The light curve analysis demonstrated that the light variation of the system is caused by eclipses combined with the effect of magnetic activity existing on the primary component. Considering the contact times derived from the theoretical synthetic light curve, we removed the eclipses from all the V band light curves, and we obtained the remaining curve for each season. Then, in order to reveal the magnetic activity behaviour along the years, we investigated the data at out-of-eclipses firstly for the short-term and secondly for the long-term variations.

To reveal the spot configuration on the primary component's surface (especially the longitudinal distributions of the spotted areas) in each short-term interval, we modelled the remaining curves under some assumptions with using the SPOTMODEL program \citep{Rib03, Rib02}. Although the data obtained in this study contain multi-band observations, the available data in the literature do not. All the data taken from both the NSVS and ASAS databases and also from both \citet{Ber06} and \citet{Fra07} contain just the V band observations. However, modelling progress in the SPOTMODEL program requires two band observations or any spot temperature factor at least. Although the large part of the using data is mono-chromatic, it is likely that we can get a temperature factors for the spotted areas from light curve analysis of I band observations in the season 2008. According to surface temperature of the primary component, the derived temperature factors are in agreement with the temperature factors found from other analogue stars \citep{Tho08}. Considering the spot temperature factor derived from the light curve analysis, we assumed the temperature factors of the spotted areas are 0.80 in the SPOTMODEL program, and we took it as constant parameters for the models in order to just determine surface distributions of the spotted areas for each short-term interval. The longitudes, latitudes and radii of the spots were taken as adjustable parameters in the program for the model of each season.

The derived models are shown in Figure 6. The models of the data taken from the NSVS database are shown in panels a and b of Figure 6, while the ASAS data are in panels c, d. For the seasons 2005 and 2006, there are some data in both the ASAS database and \citet{Ber06} and \citet{Fra07}. These data were combined for each year. The models of these years are shown in panels e and f of Figure 6. For the season 2007, some data were obtained in this study, and also some data were taken from \citet{Fra07}. The combined data were modelled, and the result is shown in panel g of Figure 6. In panels h and i, the models of light curves obtained in the seasons 2008 and 2011. The parameters of the models are listed in Table 4. In the table, the data set, mean times of observations, mean years of observations are listed in first three columns, while the spot parameters such as longitude ($I$), latitude ($b$) and radius ($g$) are listed in the following columns for two spots, respectively.

As expected, the spot distribution on the surface is rapidly changing. The distributions are varying from a year to next one. As seen from Figure 6, there is always a spotted area on the surface of the primary star, while there are two spotted areas on its surface. Figure 7 demonstrates that the spotted areas are always separated from each other. Although there are usually about 180$^\circ$ longitudinal differences between them, they are some times getting closer to each other. As seen from Figure 7, the longitudes of the one of the spotted areas exhibit a quasi-sinusoidal variation along the years, while both of them migrate toward the decreasing longitudes. Using GraphPad Prism V5.02 software \citep{Mot07}, the quasi-sinusoidal variation was fitted by a polynomial function. To test whether the polynomial fit is statistically acceptable, we computed the probability value (hereafter p-value). The value of $\alpha$ was taken as 0.005 for the p-value, which allowed us to test whether the p-value are statistically acceptable or not \citep{Daw04}. The p-value was found to be p-value $<$ 0.00167. Considering the $\alpha$ value, this means that the result is statistically acceptable.

Apart from the short-term variations, considering the remained V band light curves without any eclipses, we examined the variations of the mean brightness, the amplitude and the deeper minimum phases of light curves ($\theta_{min}$). All of them are listed in Table 5. The variations of three parameters are shown versus the years in Figure 8. The variation of the amplitudes for the remaining curves is shown in upper panel (a). The mean brightness variation is shown in middle panel (b). As it is seen from the figures, the amplitude is varying in a sinusoidal way with a period of $\sim$8.9 years. Using GraphPad Prism V5.02 software, the variation was fitted by a polynomial fit. According to the statistical analysis, just two point, which are shown by open circles in the figure, diverged from the general trend. Apart from these point, the polynomial function is fitted the general trend with the correlation coefficient of 0.91. Moreover, The p-value was found to be p-value $<$ 0.00093. Considering the $\alpha$ value, p-value the result is statistically acceptable. However, the mean brightness is dramatically decreasing through the years from 1999 to 2011. In these years, it was decreased from $10^{m}.40$ to $10^{m}.60$. Both figures indicate that the primary star has high level the magnetic activity. The distribution of the deeper minimum phases of the remaining curves is shown in the bottom panel of Figure 8.

As seen from Figure 8, the deeper minimum was located in the phase interval between $0^{P}.35$ and $0^{P}.55$ from 1999 to 2005. However, it was suddenly shifted to the phase interval between $0^{P}.95$ and $0^{P}.05$ from 2006 to 2008. It was again seen in the phases between $0^{P}.35$ and $0^{P}.355$ in the season 2011. The phenomenon is generally "flip-flop" in the literature \citep{Berd06, Ola06, Kor05, Kor07}. Considering the minima seen in the phase interval between $0^{P}.35$ and $0^{P}.55$ reveal that the minima slowly migrated toward the decreasing phases.

\section{Results and Discussion}

In this study, we tried to reveal the nature of close binary system GSC 02038-00293. The light curve analysis indicated that the mass ratio of the system is 0.35. We estimated that the mean radii are 0.87 $R_{\odot}$ for the primary and 0.27 $R_{\odot}$ for the secondary component. \citet{Kor10} had indicated that the effective temperature is 4750 $K$. They found to be $R \times \sin i = 0.88$ $R_{\odot}$. In this study, the inclination ($i$) of the system was found to be 77.91$^\circ$. In this case, the primary component's radius estimated in this study is in agreement with one given by \citet{Kor10}. On the other hand, the radius of the primary star is actually larger than an expected value. As it is seen from Figure 5, the secondary component is located almost on the ZAMS. This indicates that this component should be very young. However, it seems that the primary component has departed from the ZAMS. It is more likely that, considering their masses, the primary component should be evolved more rapidly than the secondary component. On the other hand, as it was discussed by \citet{Lop07, Mor08, Cas08, Fer09, Mor10, Tor10} and \citet{Kra11}, the case of larger radius is a common phenomena for the active stars. \citet{Lop07} demonstrated that the case is very common especially for the magnetically active stars in the mass range from 0.35 $M_{\odot}$ to 0.70 $M_{\odot}$. The mass of the primary star of the system is close to this interval. Consequently, it is more possible that its radius was found larger due to the magnetic activity. The radius of the secondary component is in agreement with the expected value for a star with 0.25 $M_{\odot}$.

The light curve analysis of I band observations of the season 2008 and the spot models of the light curves at out-of-eclipses confirmed that there are two spotted areas on this component. The models and the examining of variations at out-of-eclipses generally reveal some properties for the primary component's magnetic activity behaviour. (1) There are two active longitudes on the primary component. (2) One of them is usually active, while second one can be sometime less active. (3) Although there are two stable active longitudes on the star, the locations of the spotted areas on this star can rapidly change. Both active longitudes migrate toward the decreasing longitude. In addition, one of them exhibits a quasi-sinusoidal variation during the migration. Here, it should be noted that we assumed that the temperature factors of the spots are stable and constant along the years. Because of this assumption, the radii and latitudes of the spotted areas obtained from the spot models can clearly change with taking another temperature factor for the spots. However, the longitudes will not change. Although the temperature factor, radius and latitude depend on each other, but the longitude does not depend directly on them.

Examinations of the long-term amplitude variation of the light curves at out-of-eclipses indicate that the amplitude is varying in a sinusoidal way, while the mean brightness of the system is dramatically decreasing. In this case, it is possible that the spots should cover more part of the surface of the active component, while the spotted areas sometimes gather toward an active longitude. On the other hand, it is also possible that there can be another scenario. The decreasing of the mean brightness can be due to the any spotted polar-cap areas. The increasing total area covered by spots on the polar can cause the same effect on the mean brightness. However, the phases of the deeper minima at out-of-eclipses migrate toward the decreasing phases. The other point is that the second active longitude is more active than the first one between the seasons 2006 and 2008. This is a small clue for the "flip-flop" behaviour, which is general property seen in many active stars \citep{Berd06, Ola06, Kor05, Kor07}.

In this study, the nature of the system is a bit cleared. As a result of the analyses, GSC 02038-00293 is a close binary, whose primary component exhibits high level magnetic activity. The long-term photometric observations together with the spectral study will reveal the nature in a better view, as well.

\section*{Acknowledgments} The authors acknowledge generous allotments of observing time at the Ege University Observatory (EUO). We wish to thank Dr. Bernhard, who shared all his data with us, and braced us with delivering all the ROTSE 1 and ASAS 3 data of GSC 02038-00293. We also thank the referee for useful comments that have contributed to the improvement of the paper.

\clearpage

\begin{figure*}[h]
\hspace{0.65cm}
\includegraphics[scale=0.92, angle=0]{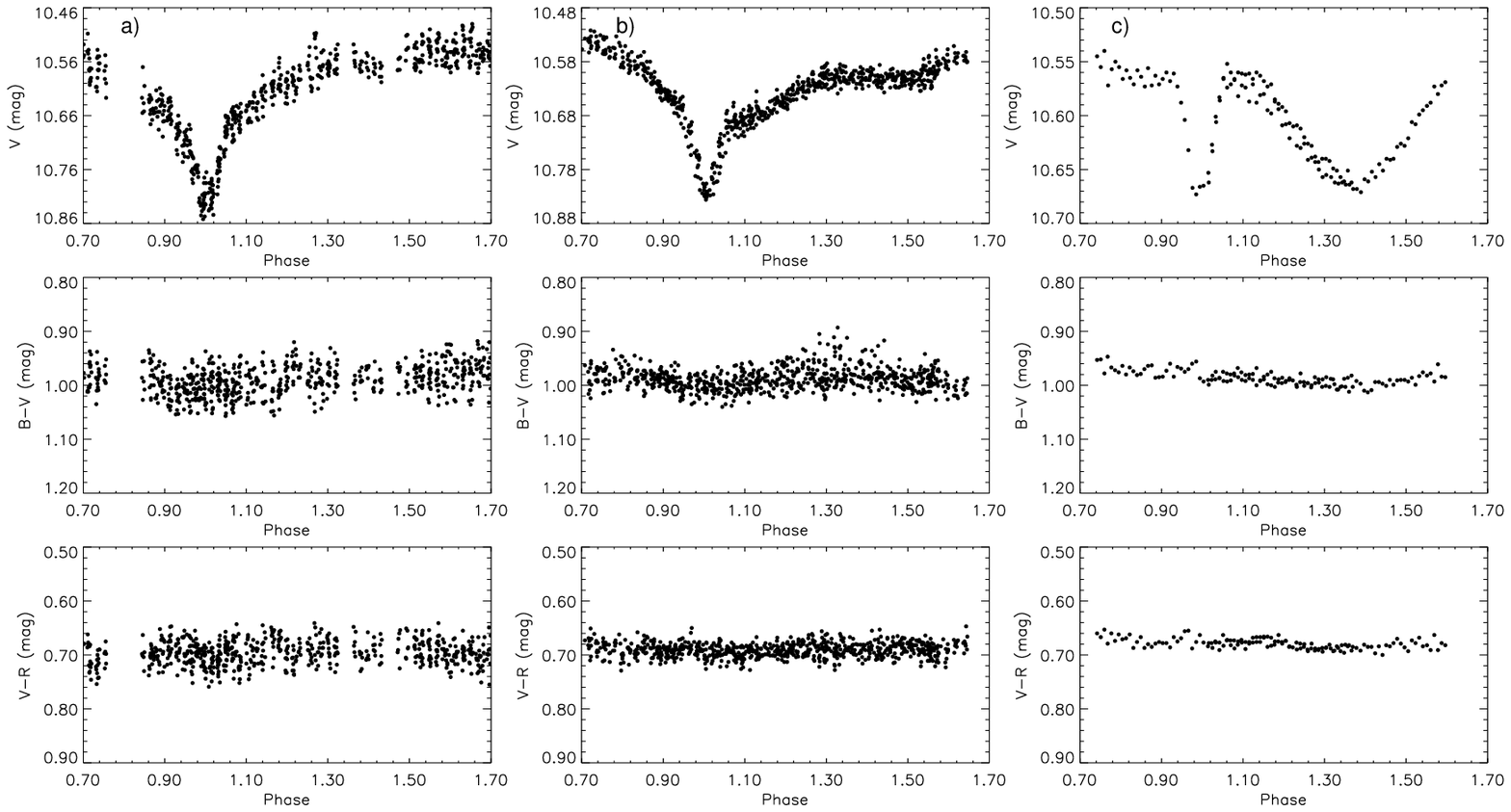}
\vspace{0.3cm}
\caption{The light and colour curves of GSC 02038-00293 for the observing seasons 2007 (a), 2008 (b) and 2011 (c).}
\label{Fig1}
\end{figure*}

\begin{figure*}[h]
\hspace{0.8cm}
\includegraphics[scale=0.92, angle=0]{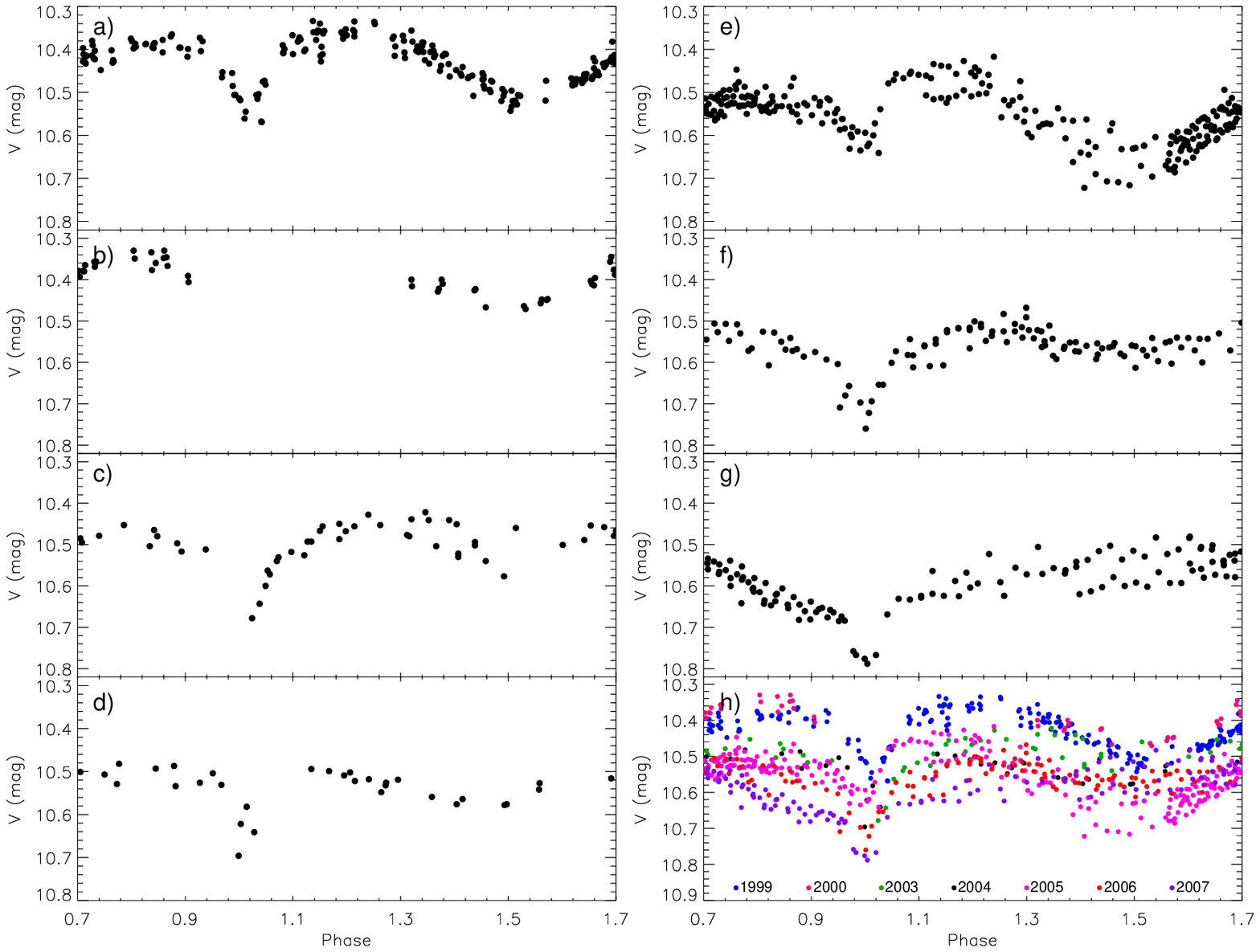}
\vspace{0.3cm}
\caption{The available observations of GSC 02038-00293 in the literature. In panels (a) and (b), the light curves of the ROTSE 1 V band data are shown for the seasons 1999 and 2000. The V band light curves of 2003 (c), 2004 (d), 2005 (e), 2006 (f) and 2007 (g), whose data were taken from the ASAS database \citep{Poj97} and from \citet{Ber06} and \citet{Fra07}. For the seasons 2005 and 2006, there are some data in both the ASAS database and \citet{Ber06} and \citet{Fra07}. To easy compare, all the data set are shown together in panel h.}
\label{Fig2}
\end{figure*}

\begin{figure*}[h]
\hspace{3.5cm}
\includegraphics[scale=0.70, angle=0]{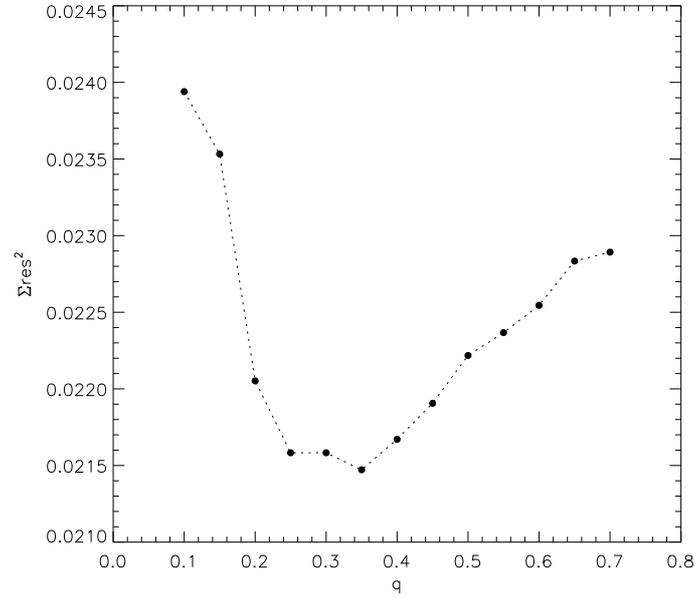}
\vspace{0.3cm}
\caption{The variation of the sum of weighted squared residuals versus mass ratio in the "q search".}
\label{Fig3}
\end{figure*}

\begin{figure*}[h]
\hspace{3.0cm}
\includegraphics[scale=0.80, angle=0]{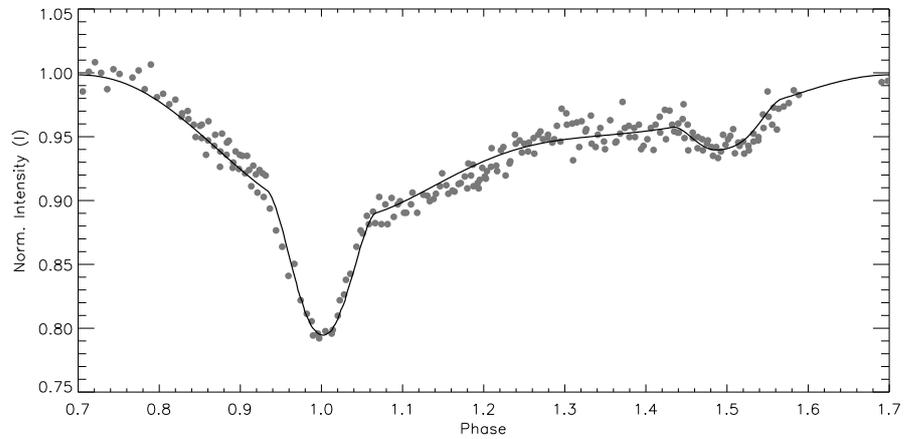}
\vspace{0.3cm}
\caption{The synthetic light curve obtained from the light curve analysis of I band.}
\label{Fig4}
\end{figure*}

\begin{figure*}[h]
\hspace{5.0cm}
\includegraphics[scale=0.80, angle=0]{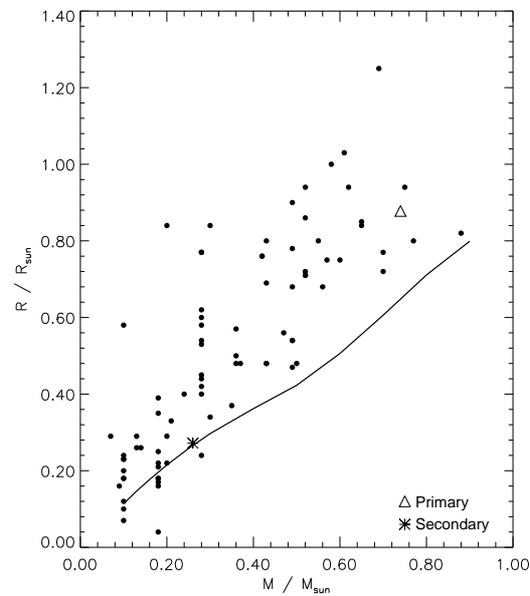}
\vspace{0.3cm}
\caption{The places of the components of GSC 02038-00293 among well-known active stars in the Mass-Radius distribution. In the figure, the filled circles represent the active stars listed in the catalogue of \citet{Ger99}. The asterisk represents the secondary component, while the open triangle represents the primary component of GSC 02038-00293. The line represents the ZAMS theoretical model developed by \citet{Sie00}.}
\label{Fig5}
\end{figure*}

\begin{figure*}[h]
\hspace{2.3cm}
\includegraphics[scale=0.8, angle=0]{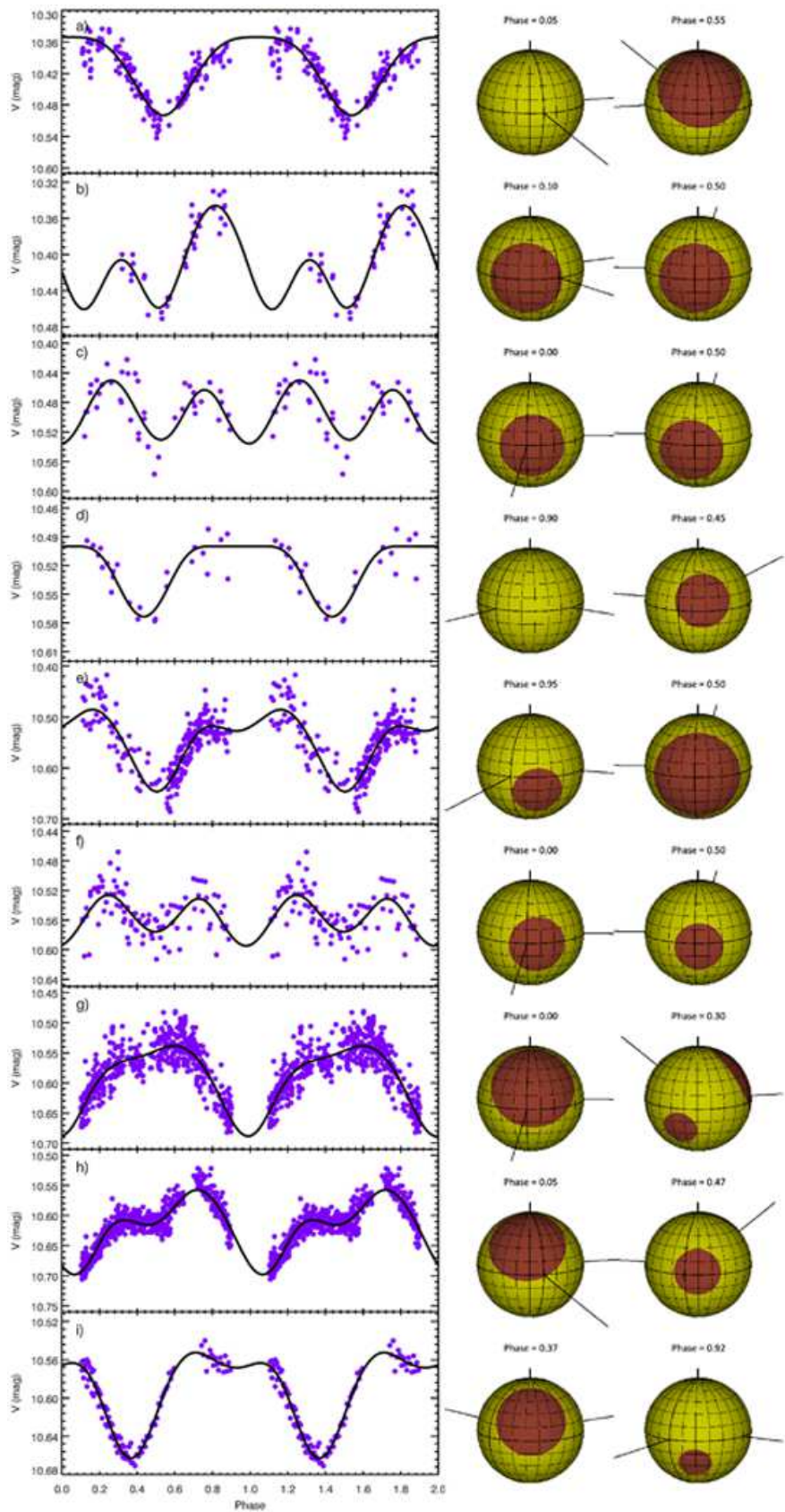}
\vspace{0.3cm}
\caption{The synthetic light curves at out-of-eclipses and 3D surface models. In each panel on left side, the filled circles represent the observations, while the lines represent theoretical fit derived by the SPOTMODEL program. The 3D surface model for two phases (especially the phases the spots seen) is seen just on right side of the light curve of each model.}
\label{Fig6}
\end{figure*}

\begin{figure*}[h]
\hspace{3.2cm}
\includegraphics[scale=0.75, angle=0]{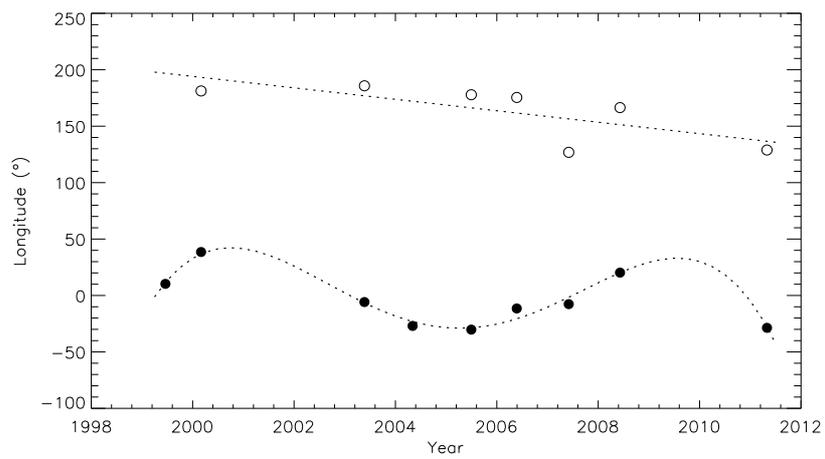}
\vspace{0.3cm}
\caption{The variation of the longitudes of the spotted areas. In the figure, filled circles represent the Spot I, while the open circles represent Spot II. The dashed lines represent the linear fit for the Spot I and the polynomial one for the Spot II.}
\label{Fig7}
\end{figure*}

\begin{figure*}[h]
\hspace{4.5cm}
\includegraphics[scale=1.00, angle=0]{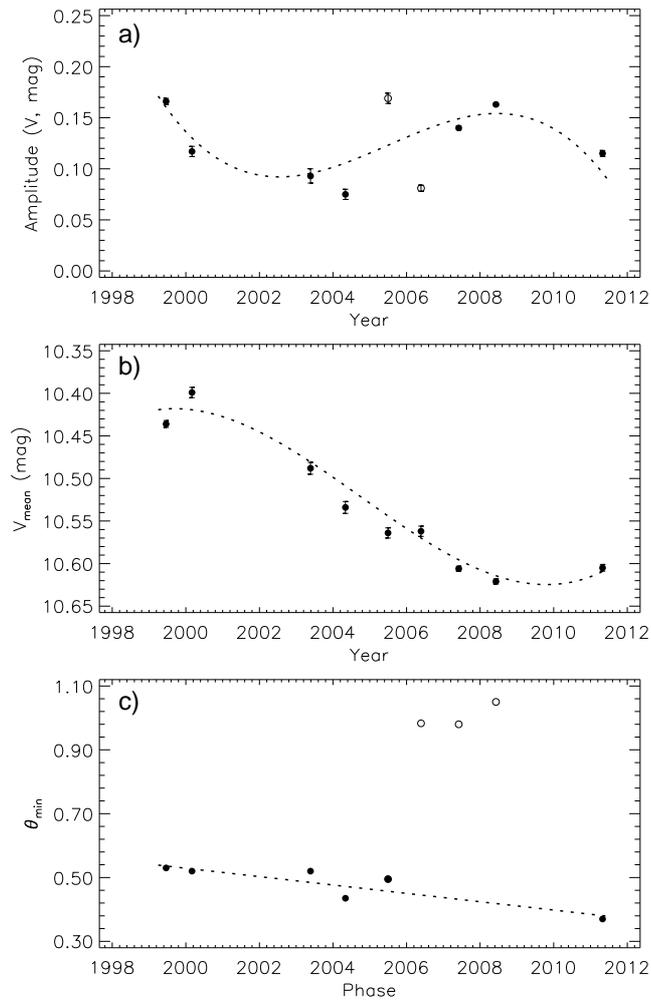}
\vspace{0.3cm}
\caption{The variations of some parameters determined from the remaining curves. a) The amplitude variation. b) The mean brightness variation. c) The spot minimum phase distribution. The filled circles represent the determined parameters in the figure, while the dashed lines represent the polynomial fits in panel a and b, while the dashed line represents the linear fit for the theta min of one spotted area. The open circles in panel a represent the points diverged from the general trend.}
\label{Fig8}
\end{figure*}

\clearpage

\begin{table*}
\begin{center}
\caption{Basic parameters for the observed stars. V band brightness and B-V index were obtained in this study.\label{tbl-1}}
\begin{tabular}{@{}lccc@{}}
\hline\hline
Star	&	RA / DE (J2000)	&	V	&	B-V	\\
Name	&	($^{h}$ $^{m}$ $^{s}$) / ($^\circ$ $^{\prime}$ $^{\prime\prime}$)	&	(mag)	&	(mag)	\\
\hline							
GSC 02038-00293	&	16 02 48.54 +25 20 38.9	&	10.540	&	0.972	\\
GSC 02038-00867 (Comparison)	&	16 05 35.85 +25 16 59.6	&	9.185	&	1.243	\\
GSC 02038-00565 (Check)	&	16 02 53.98 +25 10 43.2	&	11.811	&	0.394	\\
GSC 02038-00663 (Check)	&	16 03 13.37 +25 12 10.9	&	11.513	&	0.738	\\
\hline
\end{tabular}
\end{center}
\end{table*}

\begin{table*}
\begin{center}
\caption{The minima times of GSC 02038-00293.\label{tbl-2}}
\begin{tabular}{@{}lccc@{}}
\hline\hline
HJD (+24 00000)	&	Error	&	Type	&	Filter	\\
\hline							
54213.4434	&	0.0014	&	I	&	BVR	\\
54219.3903	&	0.0032	&	I	&	BVR	\\
54226.3264	&	0.0014	&	I	&	BVR	\\
54526.5460	&	0.0008	&	I	&	BVR	\\
54587.4837	&	0.0007	&	I	&	BVR	\\
54621.4215	&	0.0019	&	II	&	I	\\
54651.3956	&	0.0067	&	I	&	BVRI	\\
54659.3202	&	0.0038	&	I	&	BVRI	\\
54671.4488	&	0.0097	&	II	&	BVRI	\\
55675.4011	&	0.0005	&	I	&	BVRI	\\
\hline
\end{tabular}
\end{center}
\end{table*}

\begin{table*}
\begin{center}
\caption{The parameters obtained from I band light curve analysis.\label{tbl-3}}
\begin{tabular}{@{}lr@{}}
\hline\hline
Parameter	&	Value	\\
\hline			
$q$	&	0.35	\\
$i$ ($^\circ$)	&	77.91$\pm$0.87	\\
$T_{1}$ (K)	&	4750 (Fixed)	\\
$T_{2}$ (K)	&	3515$\pm$61	\\
$\Omega_{1}$	&	3.42$\pm$0.05	\\
$\Omega_{2}$	&	4.72$\pm$0.14	\\
$L_{1}/L_{T}$ (I)	&	0.977$\pm$0.067	\\
$g_{1}$, $g_{2}$	&	0.32, 0.32	\\
$A_{1}$, $A_{2}$	&	0.5, 0.5 \\
$x_{1,bol}$, $x_{2,bol}$	&	0.625, 0.625 \\
$x_{1,I}$, $x_{2,I}$	&	0.681, 0.681 \\
$<r_{1}>$	&	0.334$\pm$0.006	\\
$<r_{2}>$	&	0.103$\pm$0.004	\\
$Co-Lat_{Spot~I}$ ($^{\circ}$)	&	90.00 (fixed)	\\
$Long_{Spot~I}$ ($^{\circ}$)	&	0.00 (fixed)	\\
$R_{Spot~I}$ ($^{\circ}$)	&	54.43 (fixed)	\\
$T_{eff,~Spot~I}$	&	0.97 (fixed)	\\
$Co-Lat_{Spot~II}$ ($^{\circ}$)	&	90.00 (fixed)	\\
$Long_{Spot~II}$ ($^{\circ}$)	&	263.56 (fixed)	\\
$R_{Spot~II}$ ($^{\circ}$)	&	54.43 (fixed)	\\
$T_{eff,~Spot~II}$	&	0.98 (fixed)	\\
\hline
\end{tabular}
\end{center}
\end{table*}

\begin{table*}
\begin{center}
\caption{The spot parameters derived by the SPOTMODEL program are listed for each data set. In the table, the subscripts 1 and 2 represent Spot I and Spot II.\label{tbl-4}}
\begin{tabular}{@{}cccccccccc@{}}
\hline\hline
Data	&	Mean HJD	&	Mean	&	$l_{1}$	&	$l_{2}$	&	$b_{1}$	&	$b_{2}$	&	$g_{1}$	&	$g_{2}$	&	Data	\\
Set	&	(+24 50000)	&	Year	&	($^\circ$)	&	($^\circ$)	&	($^\circ$)	&	($^\circ$)	&	($^\circ$)	&	($^\circ$)	&	Source	\\
\hline
A	&	1350.2176	&	1999.46	&	10	&	-	&	144	&	-	&	53	&	-	&	1	\\
B	&	1607.0397	&	2000.17	&	39	&	-	&	358	&	-	&	43	&	-	&	1	\\
B	&	1607.0397	&	2000.17	&	-	&	181	&		&	358	&	-	&	43	&	1	\\
C	&	2784.3294	&	2003.39	&	354	&	-	&	357	&	-	&	38	&	-	&	2	\\
C	&	2784.3294	&	2003.39	&	-	&	186	&		&	351	&	-	&	37	&	2	\\
D	&	3131.0668	&	2004.34	&	333	&	-	&	168	&	-	&	31	&	-	&	2	\\
E	&	3553.9587	&	2005.50	&	330	&	-	&	342	&	-	&	28	&	-	&	2, 3	\\
E	&	3553.9587	&	2005.50	&	-	&	178	&		&	357	&	-	&	55	&	2, 3	\\
F	&	3881.8078	&	2006.39	&	349	&	-	&	357	&	-	&	32	&	-	&	2, 3	\\
F	&	3881.8078	&	2006.39	&	-	&	175	&		&	355	&	-	&	28	&	2, 3	\\
G	&	4255.8110	&	2007.42	&	352	&	-	&	30	&	-	&	51	&	-	&	3, 4	\\
G	&	4255.8110	&	2007.42	&	-	&	127	&		&	24	&	-	&	20	&	3, 4	\\
H	&	4624.8493	&	2008.43	&	20	&	-	&	40	&	-	&	47	&	-	&	4	\\
H	&	4624.8493	&	2008.43	&	-	&	166	&		&	3	&	-	&	26	&	4	\\
I	&	5683.9083	&	2011.33	&	331	&	-	&	332	&	-	&	18	&	-	&	4	\\
I	&	5683.9083	&	2011.33	&	-	&	129	&		&	24	&	-	&	41	&	4	\\
\hline 
\end{tabular} 
\end{center} 
$^{1}$ The NSVS Database \citep{Woz04} \\
$^{2}$ The ASAS Database \citep{Poj97} \\
$^{3}$ \citet{Ber06} and \citet{Fra07} \\
$^{4}$ This Study \\
\end{table*}

\begin{table*}
\begin{center}
\caption{ Some parameters determined from the remaining V light curves.\label{tbl-5}}
\begin{tabular}{@{}ccccccccc@{}}
\hline\hline
Data	&	Mean HJD	&	Mean	&	$\theta_{min}$	&	$V_{min}$	&	$V_{max}$	&	Amplitude	&	$V_{mean}$	&	Data	\\
Set	&	(+24 50000)	&	Year	&	&	(mag)	&	(mag)	&	(mag)	&	(mag)	&	Source	\\
\hline
A	&	1350.2176	&	1999.46	&	0.530	&	10.519	&	10.353	&	0.166	&	10.436	&	1	\\
B	&	1607.0397	&	2000.17	&	0.520	&	10.457	&	10.340	&	0.117	&	10.399	&	1	\\
C	&	2784.3294	&	2003.39	&	0.520	&	10.534	&	10.441	&	0.093	&	10.488	&	2	\\
D	&	3131.0668	&	2004.34	&	0.435	&	10.571	&	10.496	&	0.075	&	10.534	&	2	\\
E	&	3553.9587	&	2005.50	&	0.495	&	10.648	&	10.479	&	0.169	&	10.564	&	2, 3	\\
F	&	3881.8078	&	2006.39	&	0.983	&	10.602	&	10.521	&	0.081	&	10.562	&	2, 3	\\
G	&	4255.8110	&	2007.42	&	0.980	&	10.676	&	10.536	&	0.140	&	10.606	&	3, 4	\\
H	&	4624.8493	&	2008.43	&	1.050	&	10.702	&	10.539	&	0.163	&	10.621	&	4	\\
I	&	5683.9083	&	2011.33	&	0.370	&	10.662	&	10.547	&	0.115	&	10.605	&	4	\\
\hline 
\end{tabular} 
\end{center} 
$^{1}$ The NSVS Database \citep{Woz04} \\
$^{2}$ The ASAS Database \citep{Poj97} \\
$^{3}$ \citet{Ber06} and \citet{Fra07} \\
$^{4}$ This Study \\
\end{table*}


\end{document}